\documentclass[reprint,footnoteinbib,prl,twocolumns]{revtex4-1}
\usepackage{amsfonts,amsmath,amssymb}
\usepackage[pdftex]{graphicx}
\usepackage{mathrsfs}
\usepackage{wasysym}
\usepackage{enumerate}
\usepackage{bm}

\usepackage[svgnames]{xcolor}
\usepackage{braket}
\usepackage[
    colorlinks=True,linkcolor=DarkRed,citecolor=ForestGreen,urlcolor=MediumBlue,
    pdfstartview=FitH,bookmarks=False,pdfpagemode=UseNone
]{hyperref}

\begin{document}

\title{High resolution spectroscopy and narrow resonances from InGaN quantum dots in GaN nanowires.}

\author{Cameron Nelson$^{1,2,*}$, Saniya Deshpande$^{1,2}$, Albert Liu$^{1,2}$, \\ Shafat Jahangir$^{1,2}$, Pallab Bhattacharya$^{1,2}$, Duncan G. Steel$^{1,2,3,*}$ \\
\normalsize{$^1$Center for Photonics and Multiscale Materials, University of Michigan, Ann Arbor, Michigan} \\
\normalsize{$^2$ Department of Electrical Engineering and Computer Science, University of Michigan, Ann Arbor, Michigan} \\
\normalsize{$^3$H.M. Randall Laboratory of Physics, University of Michigan, Ann Arbor, Michigan}
}

\begin{abstract}
High resolution coherent nonlinear optical spectroscopy of an ensemble of red-emitting InGaN quantum dots in GaN nanowires is reported.  The data show a pronounced atom-like interaction between resonant laser fields and quantum dot excitons at low temperature that is difficult to observe in the linear absorption spectrum due to inhomogeneous broadening from indium fluctuation effects. We find that the nonlinear signal persists strongly at room temperature. The robust atom-like room temperature response indicates the possibility that this material could serve as the platform for proposed excitonic based applications without the need of cryogenics.
\end{abstract}

\maketitle

%%%%%%%%%%% Eight-level diagram figure with transitions
\begin{figure}
 \includegraphics[width=0.7\linewidth]{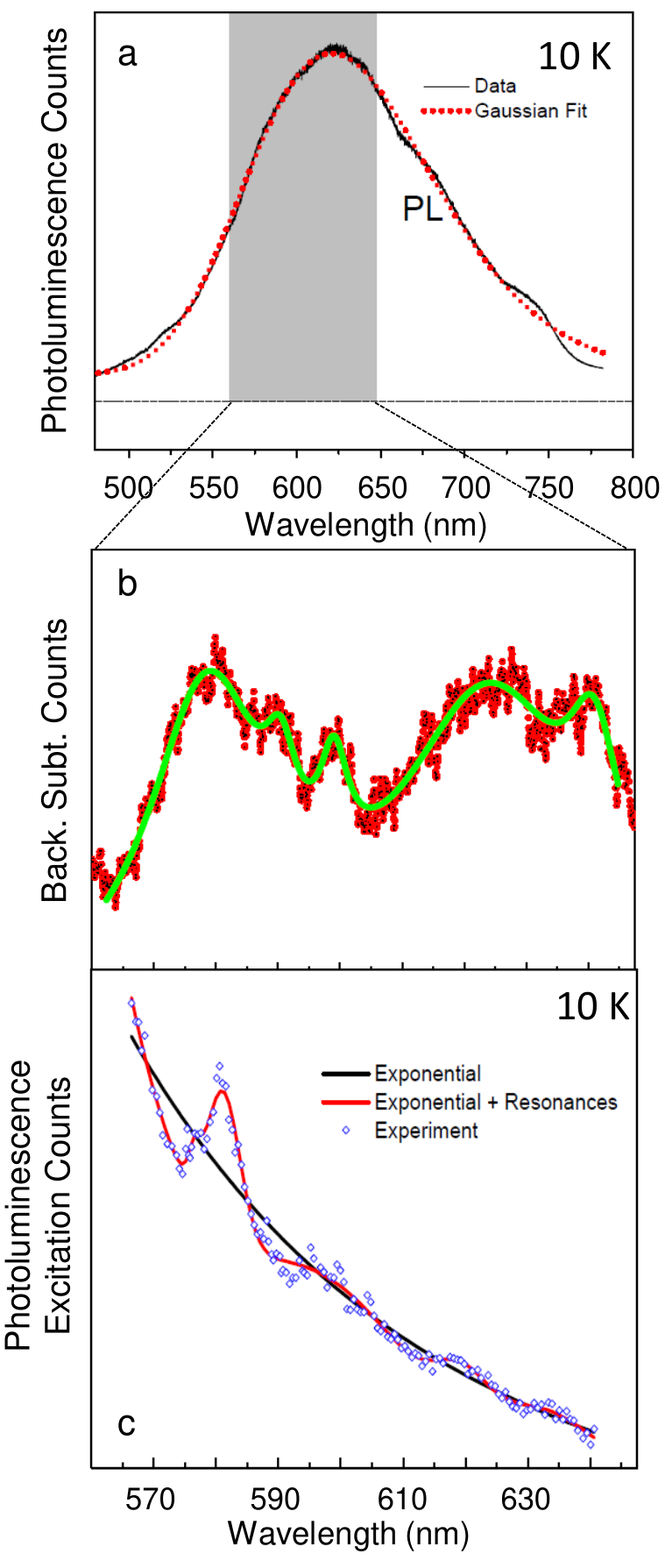}
  \caption{a. Photoluminescence spectrum of $\sim~$10$^3$ DINWs using a 405 nm wavelength laser for excitation measured at 10 K. A Gaussian fit to the data is shown by the red lines.  b. Photoluminescence spectrum after subtraction of Gaussian reveals discrete resonances. c. Photoluminescence excitation spectrum (blue points) with exponential fit shown as the solid black line and the exponential fit including excitonic resonances shown in the solid red line. The resonances are absorption peaks from individual DINWs or small groups of DINWs.}
 \label{fig:PL}
\end{figure}
%%%%%%%%%%%%%

Recent studies have shown single photon emission behavior in III-Nitride quantum-confined structures up to room temperature \cite{doi:10.1021/nl404400d,DeshpandeFrostHazariEtAl2014}. This discovery provides the basis for future research efforts toward room temperature coherent control of the III-Nitride quantum dot exciton states \cite{BonadeoErlandGammonEtAl1998,StievaterLiSteelEtAl2001,KamadaGotohTemmyoEtAl2001} or spin ground states \cite{DuttChengLiEtAl2005,SantoriTamaratNeumannEtAl2006,
PressLaddZhangEtAl2008} in III-Nitride quantum dots charged with a single electron for various applications including quantum information processing. In recent studies at low temperature, the population of a III-Nitride quantum dot excited state was coherently manipulated using laser pulses to demonstrate evidence of Rabi rotations \cite{HolmesKakoChoiEtAl2013,ReidKocherZhuEtAl2014}, which provides a proof-of-concept for qubit manipulations in InGaN quantum information processing. Red emitting InGaN quantum dots have become particularly attractive recently for quantum information applications due to their larger quantum confinement compared to green-blue emitting dots and the higher sensitivity of Si-based photo detectors at these wavelengths \cite{Hadfield2009}.  Development of this materials for applications such as optically controlled switching or quantum information processing requires a full understanding of the linear and nonlinear optical response and the associated relaxation and spectral lineshapes \cite{BonadeoChenGammonEtAl1998,WuLiSteelEtAl2004}.  These issues are complicated in this material by the presence of indium disorder effects \cite{MartinMiddletonOa€™DonnellEtAl1999,SchoemigHalmForchelEtAl2004,
WooBugnetNguyenEtAl2015,LiFischerWeiEtAl2010,
SchulzCaroCoughlanEtAl2015,ChichibuUedonoOnumaEtAl2006} that can complicate many details of this response as well as leading to inhomogeneous broadening and spectral diffusion.  

In this Letter we use high resolution coherent nonlinear optical spectroscopy on In$_{0.54}$Ga$_{0.46}$N dots-in-nanowires (DINWs) at low temperature to reveal pronounced resonances that are not easily seen in the photoluminescence or photoluminescence excitation (PLE) spectrum. We infer that the resonances are caused by nonlinear absorption of quantum dot excitons.  The exciton resonances are not easily observed in the linear absorption spectrum due to absorption from a distribution of background states, likely formed by indium disorder effects. Ultra-narrow resonances in the third order ($\chi^{(3)}$) coherent nonlinear optical spectrum show the excitons are coupled to long-lived metastable background states that lead to strong saturation of the quantum dot exciton optical absorption. Additional data demonstrates that the nonlinear optical resonances are clearly visible at room temperature with line widths similar to the low temperature data. This data provides the first evidence of resonant optical absorption in quantum confined III-N excitons at room temperature, a promising result for on-chip quantum information research.

To the best of our knowledge, this is the first report that the optical absorption below the band edge in DINWs can arise from two separate mechanisms:  (1) the optical absorption associated with the usual enhanced optically induced dipole due to the strong Coulomb coupling between electron and hole that gives rise to the exciton; and (2) optically active background states formed by material disorder that form a quasi-continuum and energy states and hence absorption. Our data show an \textit{increase in the absorption} in the presence of optical excitation, of importance to such devices as optical limiters \cite{StrylandWuHaganEtAl1988}. The nonlinear signal can be used to distinguish between the two absorption mechanisms: the quantum confined exciton states show saturation of their optical absorption, while the background states show increased absorption. Degenerate-four-wave-mixing measurements in InGaN/GaN quantum wells has revealed a similar enhancement in the nonlinear signal of exciton resonances compared to a featureless linear absorption background caused by disorder states \cite{KundysWellsAndreevEtAl2006}. 

The self-assembled GaN/In$_{0.54}$Ga$_{0.46}$N nanowire/DINW sample featured in this study was grown along the c-axis in the wurtzite configuration using plasma-assisted MBE on (001) silicon. Details of the growth procedure can be found in Refs. \cite{GuoZhangBanerjeeEtAl2010,JahangirMandlStrassburgEtAl2013,
DeshpandeFrostYanEtAl2015}. The sample emission wavelength is $\lambda$ $\sim$ 620 nm. The nanowires in this study are $\sim$30 nm in diameter with a 3 nm thick InGaN active region and each nanowire contains 8 DINWs.  Evidence of self-formed InN-rich quantum dots in the center of the disk from similarly grown samples has been found from low temperature photoluminescence (PL) and transmission electron microscopy (TEM) studies \cite{DeshpandeFrostYanEtAl2015,TourbotBougerolGrenierEtAl2011,
TourbotBougerolGlasEtAl2012,ChangWangLiEtAl2010}.  Isolated DINWs from similar samples show evidence of single photon emission up to 30 K \cite{DeshpandeFrostYanEtAl2015} from second order correlation measurements. Following MBE growth, the wires are removed from the silicon and dispersed on a sapphire substrate at an areal density of$\sim$1x10$^{11}$ cm$^{-2}$.

The 10 K high resolution ensemble PL (1.5 $\mu$m$^2$ excitation laser spot size) of the sample shows a broad Gaussian line shape (FWHM $\sim$300 meV ($\sim$100 nm) as shown in Fig. 1a).  Superimposed on the Gaussian are narrow features shown in Fig. 1b after removal of the broad Gaussian background (FWHM$\sim$10-30 meV ($\sim$3-10 nm)). In the PL spectrum we observe a variation in the emission energies of the narrow features of about 30-50 meV (10-15 nm). The PL intensity of individual resonances is linear with the excitation laser intensity (I$_{PL}$$\sim$I$_{exc}$$^{1.00\pm0.03}$) before saturating and shows no blue shift in frequency, in contrast to what is reported in InGaN quantum wells \cite{MartinMiddletonOa€™DonnellEtAl1999,SchoemigHalmForchelEtAl2004}. The optical setup is described further in the Supplemental Material \cite{Supp.Mat.}.  

The independence of the PL emission energy on excitation intensity and the width of the resonances suggest that the peaks belong to emission from DINW excitons in individual quantum DINWs or small groups of DINWs \cite{DeshpandeFrostHazariEtAl2014,DeshpandeFrostYanEtAl2015}.  The exciton emission energy in individual DINWs is determined by the InGaN disk thickness, diameter and indium concentration \cite{YanJahangirWightEtAl2015,SekiguchiKishinoKikuchi2010,
ZhangTengHillEtAl2013}, which are known to vary across the sample surface \cite{YanJahangirWightEtAl2015}. The emission energy of individual DINW excitons is a complicated function of these parameters since a variation in the DINW size affects the strain relaxation that changes the internal electric fields \cite{ZhangTengHillEtAl2013,SacconiAufderMaurDiCarlo2012,
ZhangHillTengEtAl2014} quantum confinement \cite{ZhangTengHillEtAl2013} and indium concentration gradients \cite{TourbotBougerolGlasEtAl2012,PereiraCorreiaPereiraEtAl2001,
MedhekarHegadekatteShenoy2008}. The variation in emission energies of the individual groups of DINWs observed in this sample is not unusual considering that $\sim$30 meV changes in emission energy were observed for small variations in size parameters in blue emitting DINWs \cite{ZhangHillTengEtAl2014}. The distribution of possible DINW emission energies, determined by the broad Gaussian resonance, is $\sim$10 times the width of the individual DINW peaks, which is consistent with reported results for similarly grown samples \cite{DeshpandeHeoDasEtAl2013}. 

The PLE data at the high energy side of the PL spectrum mainly consists of a relatively featureless, monotonically increasing background, which is well fit to an exponential as shown by the fit in Fig. 1c. The 10 K PLE spectrum also shows slight enhancements of signal at the same wavelengths as the weak resonances observed in PL.  The PLE data illustrates the continuum of background states that exist in this system. We attribute the background states in the PLE spectrum to states formed by indium disorder, such as the atomistic defects known to exist in bulk InGaN \cite{WooBugnetNguyenEtAl2015, SchulzCaroCoughlanEtAl2015, ChichibuUedonoOnumaEtAl2006} or possibly small InN-rich clusters \cite{MartinMiddletonOa€™DonnellEtAl1999,SchoemigHalmForchelEtAl2004}.  This data shows the first evidence that these types of disorder states exist in significant density within  DINW structures. 

\begin{figure}
 \includegraphics[width=0.9\linewidth]{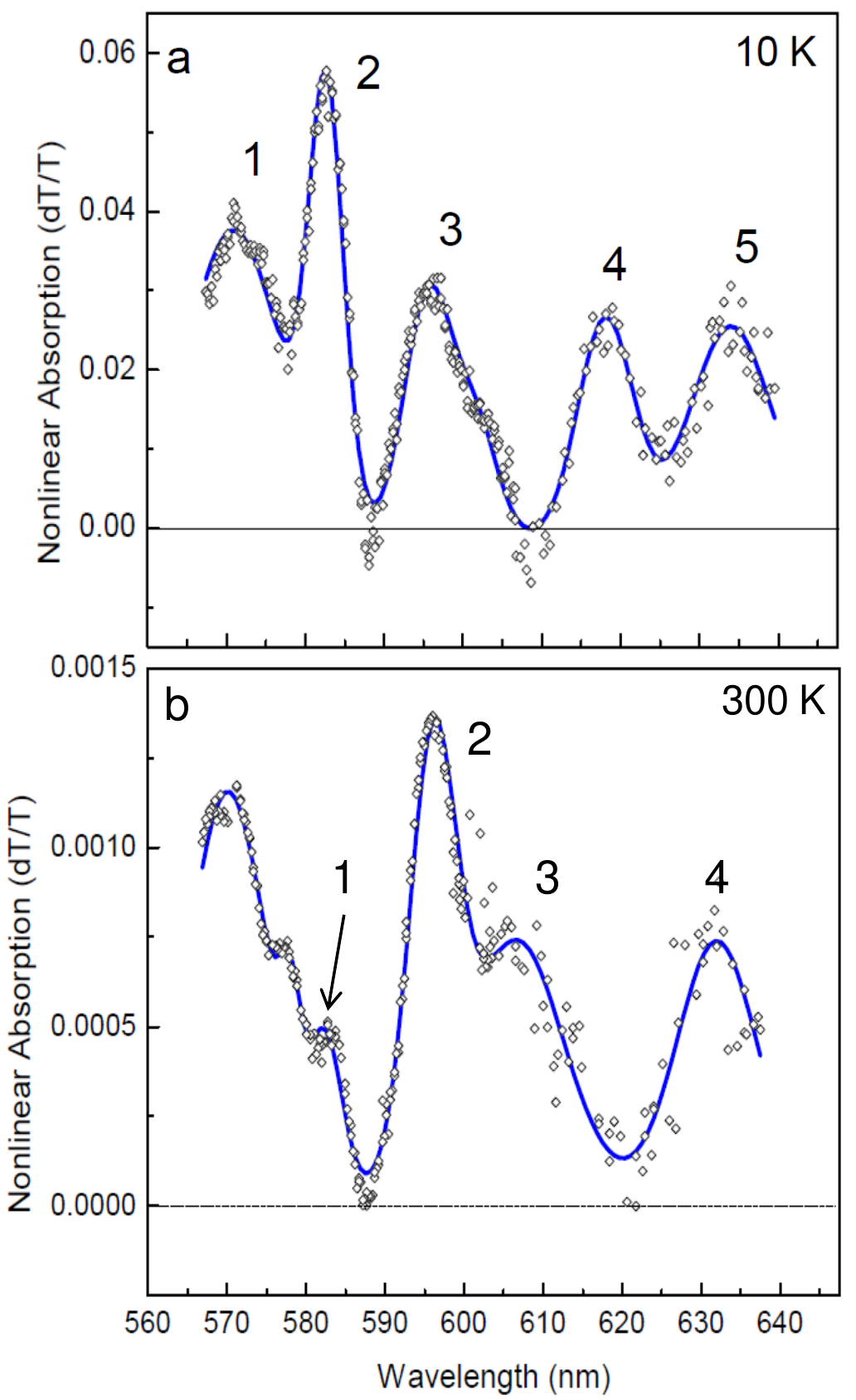}
  \caption{a. 10 K modulated absorption spectrum with constant negative dT/T offset subtracted. The peaks in the 10 K modulated absorption spectrum show good overlap with the PL after the Gaussian is subtracted. b. 300 K modulated absorption spectrum with constant negative dT/T background subtracted from the data. The peaks in Figs. 2a and 2b are fit using multiple Lorentzians with FWHMs $\sim$20-50 meV}
 \label{fig:ModAbs}
\end{figure}

 Using a 405 nm (pump) amplitude modulated laser field to excite the DINWs above the band edge and phase-sensitive-detecting the homodyne detected signal in the vicinity of the PL emission shows two primary features, as indicated above: first, the signal shows a large relatively flat negative dT/T (dT = differential transmission = T$_{Pumpon}$ - T$_{Pumpoff}$, where T is the sample transmittance) corresponding to an increase in absorption that is subtracted and not shown in Figs. 2a and 2b, (see Supplemental Material \cite{Supp.Mat.} for original data); and second, there are strong resonances seen as a relative increase in dT/T signal (meaning reduced absorption, i.e., saturation) that overlap with the weak resonances observed in the PL as shown in Fig. 1a. With the pump far to the blue of the probe field, the primary contribution to the signal is from changes in the linear absorption spectrum induced by the pump.  The magnitude of signal strength of the background and resonances, while comparable, is generally not correlated and varies depending on the area of the sample that is probed. 

It is important to note that unlike the PL (Fig. 1a) and PLE (Fig. 1c) spectra, the resonances are the dominant feature in the nonlinear absorption spectrum (except for the background offset), similar to the result obtained in Ref. \cite{KundysWellsAndreevEtAl2006}. Remarkably, the strong resonant features clearly observed in the nonlinear spectrum persist up to room temperature (Fig. 2b). The peak numbers in Fig. 2a are referenced to those in Fig. 2b by comparing the energy separation between successive peaks.  The exciton peaks at 10 K (Fig. 2a) are blue shifted compared to the 300 K exciton peaks (Fig. 2b) by about $\Delta\lambda$ = 14 nm, which is in good agreement with the shift expected from the Varshni equation \cite{Varshni1967}, which gives $\Delta\lambda$ = 13.6 nm, assuming the Varshni parameters are $\alpha$ = 0.55 and $\beta$ = 719 \cite{VurgaftmanMeyerRam-Mohan2001}. The line width of the resonances is slightly larger at room temperature compared to low temperature, however the ratio of room temperature to low temperature line widths does not exceed a factor of $\sim$2. At the same time, the strength of the nonlinear signal typically decreases by 1-2 orders of magnitude at room temperature compared to low temperature. We will see below that this could be due to a temperature dependent interaction between excitons and metastable background states. Furthermore, the resonances observed in the nonlinear absorption spectrum in this study overlap with the weak resonances observed in the PL, as shown in Fig. 1. We therefore attribute the resonances in the nonlinear absorption spectrum to quantum dot exciton states in the DINW. 

Phase modulation techniques \cite{Supp.Mat.,JamesonGrattonHall1984,PrinFlSpec} shows that the steady state modulated absorption spectrum in Fig. 2a has a very slow decay time of $\sim$10-100 $\mu$s at 10 K, depending on the specific area of sample under study, which is a clear indication that the steady-state nonlinear optical signal is dominated by processes involving background disorder states with long lifetimes. Phase modulation techniques can also be used to extract the activation energies of the background states based on the temperature dependence of the decay rate \cite{Supp.Mat.,JamesonGrattonHall1984,PrinFlSpec,SemDevDes}(Fig. 3).  The activation energy varies between $\sim$5-10 meV, due to the DINW inhomogeneity. These results are similar to previous studies on yellow-amber emitting InGaN layers that have shown that disorder states with similar activation energies may form in the InGaN layer \cite{LiFischerWeiEtAl2010}. We therefore tentatively assign the slow decay of the nonlinear signal to effects related to similar metastable states in the DINW system.  Since the strength of nonlinear optical response varies inversely with the overall excitation decay rate, the weakening of the nonlinear optical response with increasing temperature is therefore partially due to a decrease in the relaxation times of the metastable background states. We note that other effects, such as phonon interactions, will also cause a temperature dependent decrease in the nonlinear signal. 
\begin{figure}
 \includegraphics[width=0.8\linewidth]{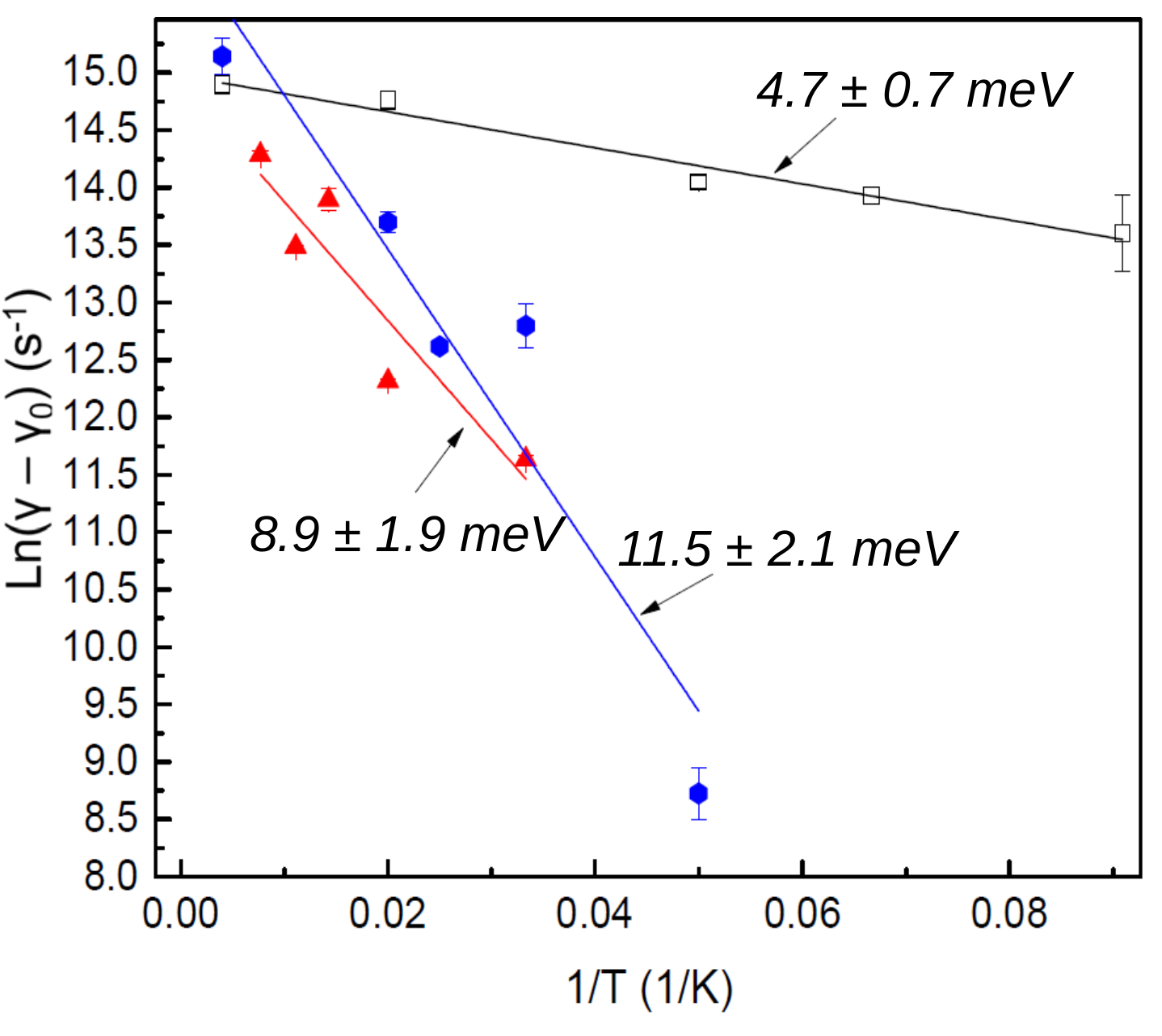}
  \caption{Arrhenius plot of decay rate of nonlinear (modulated absorption) signal ($\gamma$) subtracted by a temperature independent decay rate $\gamma_0$  for three regions of the sample at a fixed probe wavelength (617 nm) obtained using phase modulation techniques. Each colored data set represents a different (approximate) region of the sample. The error bars and decay rates vary significantly in the measurement due to sample expansion and contraction as a function of temperature. The error bars are determined by the error in the mean. The temperature independent decay rate was found for both sample regions by fitting the data Ln[$\gamma$(1/T)- $\gamma_0$] to an linear function similar to Supplemental Equation (1) (see Supplemental Material) with an offset given by $\gamma_0$, which also varied significantly across the sample but is generally $\sim$1x10$^6 s^{-1}$.}
 \label{fig:Arrhenius}
\end{figure}

 In Fig. 4a we see the nearly degenerate coherent nonlinear optical spectrum for various pump frequencies when the fixed 405nm laser is replaced with a tunable dye laser. The spectra show the surprising result that the spectrum is virtually independent of the pump frequency, even when the pump frequency is tuned below the probe frequency and for pump-probe detunings of $\sim$100 meV. Furthermore, the spectrum is virtually identical to the modulated absorption spectrum, including the negative dT/T background signal and the long ($\sim$10-100 $\mu$s) decay time of the signal. The strength of the negative dT/T background signal is also unaffected by the pump frequency over the range shown in Fig. 4a. We note that, for pump-probe detunings of $\geq$200 meV, both the positive dT/T resonances and the negative dT/T background begin to decrease significantly, however the relative strength of the positive dT/T resonances does not change. This data suggests that the nearly degenerate signal also results from optical physics related to metastable background disorder states. The discussion below will focus on this and on the various physical mechanisms that can give rise to a pump frequency-independent signal. 
\begin{figure}
 \includegraphics[width=0.9\linewidth]{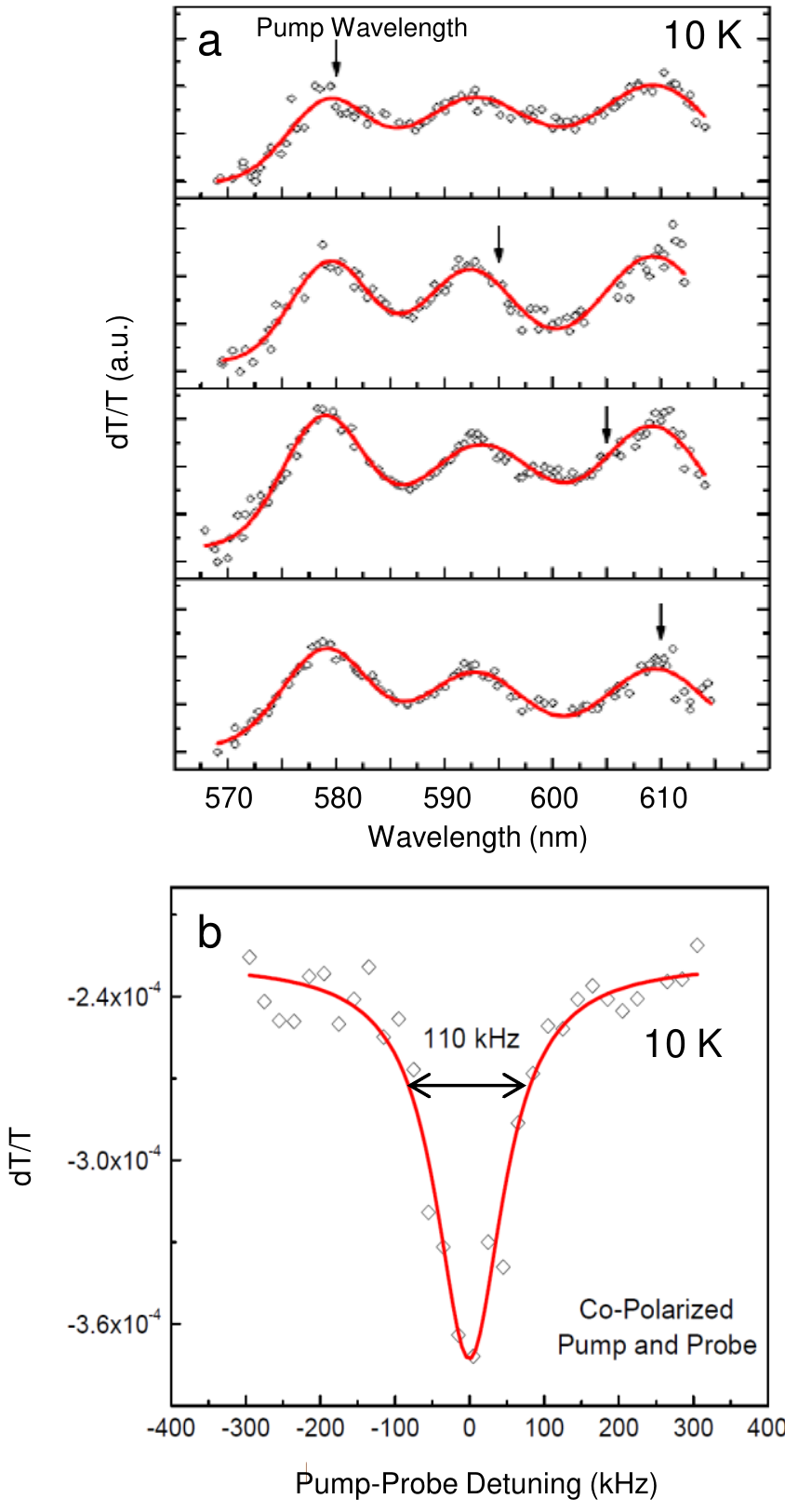}
  \caption{a. Nearly degenerate pump-probe (NDPP) data with the pump wavelength shown as the black arrow. The y axis (dT/T) is the same for all 4 panels and shows that the strength of each resonance (which are assumed to be from individual DINWs and are fit using Lorentzian lineshapes) does not depend on the pump wavelength. A constant negative dT/T background is subtracted from each plot, similar to the data shown in Fig. 2. b. Coherent population pulsation component of NDPP spectrum taken using the methods of Ref. \cite{SteelRand1985} with negative background signal included.}
 \label{fig:NddT}
\end{figure}

Modification of excited state decay by coupling to metastable states change the nonlinear optical spectrum dramatically compared to the isolated two level system \cite{Lamb1964,SteelRand1985}. In particular, the coherent coupling between the optical beams at frequencies ωpump and ωprobe, respectively, contains another spectral feature due to an additional resonant denominator \cite{Supp.Mat.,Lamb1964,SteelRand1985} arising from the presence of the metastable state.  The long life time of the metastable state results in a strong enhancement of the response that dominates the nonlinear optical spectrum over the usual population pulsation terms where the life times are determined by the excited state decay \cite{Supp.Mat.,Lamb1964}. In our sample, an ultranarrow ($\sim$110 kHz linewidth) resonance, arising from coherent population pulsations first identified by Lamb \cite{Lamb1964}, was observed as shown in Fig. 4b using the methods in Ref. \cite{SteelRand1985}. The details of this spectroscopy has been discussed extensively and reports on numerous fundamental physical parameters of the system \cite{ShenNLOpt,Berman,BoydNLOptic}.  Notice that this resonance appears as a dip in the response (which is mostly negative) rather than a positive peak as the standard theory would predict \cite{Lamb1964}. The negative dip can be attributed to the fact that the incoherent component of the third order nonlinear signal (that has both positive and negative dT/T components) is mostly negative \cite{Supp.Mat.}. By fitting the ultranarrow resonance to a Lorentzian line shape, the decay time of the metastable state is 53.5 $\pm$ 5.0 $\mu$s, in good agreement with the decay time found from phase modulation spectroscopy above (47.2 $\pm$ 1.7 $\mu$s).

In the Supplemental Material \cite{Supp.Mat.}, we consider a model in which the pump frequency independent data shown in Fig. 4a arises due to an indirect optical excitation that transfers population from the ground state of the quantum dot exciton to a disorder state \cite{Supp.Mat.}. In principle, this model provides a signal that is independent from the pump-exciton detuning and instead follows the energy distribution of background states.  However, this assumes that the exciton itself is not coupled to the metastable background states.  The data does not allow us to confirm this assumption, and if the exciton is coupled to the metastable state, we would still find a resonant enhancement when the pump was tuned within exciton resonance \cite{Supp.Mat.}.  

Another possibility is that the metastable background states are simultaneously responsible for exciton saturation in both the positive dT/T associated with the exciton resonances and the negative dT/T associated with the background.  Due to remaining strain in the InGaN DINW system, a piezoelectric field is expected to exist in the DINW, which is strongest at the center of the disk \cite{SacconiAufderMaurDiCarlo2012,ZhangHillTengEtAl2014,
PereiraCorreiaPereiraEtAl2001,WuLinHuangEtAl2009}. Because of the resulting internal electric fields and the antiparallel direction of the field with respect to the growth axis \cite{SacconiAufderMaurDiCarlo2012,ZhangHillTengEtAl2014,
PereiraCorreiaPereiraEtAl2001,WuLinHuangEtAl2009}, the electron and hole wave functions are pushed to the top and bottom of the DINW, respectively. The pump beam excites the disorder states where the relaxation from these states is characterized by a long lifetime. Electron-hole pairs that fill the metastable states are localized separately from the region of the localized exciton and can partially screen the internal piezoelectric field in the DINW. This would imply the DINW exciton should experience an increased electron-hole overlap and a blue-shift in wavelength (seen as a derivative line-shape in the dT at the level of $\chi^{(3)}$ due to the screened internal electric field via the Quantum-Confined Stark Effect (QCSE) \cite{MillerChemlaDamenEtAl1984}.  Instead we observe a pump induced decrease in absorption.  We attribute this effect to a decrease in exciton transition moment due to the exciton-metastable state interaction (see Supplemental Material \cite{Supp.Mat.}). It has been shown that charges that occupy charge trap states in close vicinity to quantum confined excitons in semiconductor nanostructures can decrease the oscillator strength of the lowest exciton transition when the exciton hole wavefunction becomes localized in the vicinity of the trap electron due to the large hole effective mass compared to the electron effective mass \cite{WangSunaMcHughEtAl1990,HilinskiLucasWang1988}. The quantum dot exciton states may be especially susceptible to this type of interaction, as it is known from nano-cathodoluminescence measurements that the largest concentration of nonradiative localized states (e.g. traps) occurs at the bottom of the DINW \cite{TourbotBougerolGlasEtAl2012}, in the same vicinity as the hole wave function.  

As discussed above, the negative dT/T offset signal appears to be mostly constant as a function of probe frequency over the spectral window featured in the data (Fig. 2 and Supplemental Material \cite{Supp.Mat.}).   If the background disorder states behave as a continuum, because of band gap renormalization there is a red-shift of the continuum states \cite{HaugSchmitt-Rink1984,ChemlaMiller1985} which would lead to a red-shift at second order in the excitation field of the PLE absorption spectrum.  This then leads to a negative and constant dT/T offset when measured in the $\chi^{(3)}$ limit.   Alternatively, it is possible that the background states experience a combination of blue shifting and an increase in electron hole overlap (negative dT/T) from the reverse QCSE as the internal piezoelectric field of the DINW is screened by the metastable states. Detailed understanding of the QCSE in this system is difficult since both the strain relaxation and InN concentration is not uniform. 

Finally, we note that the experimental approach used in this paper often shows evidence of spectral hole burning because of inhomogeneous broadening.  The absence of spectral hole burning can lead to the conclusion that the measured linewidths (outside the resonance associated with population pulsations) measures the dipole dephasing rate (T$_2^{-1}$).  The models in this paper in fact show that even with inhomogeneous broadening, no hole burning is expected even with long decoherence times at the limit of $\chi^{(3)}$ but will be evident in higher order spectroscopy.

This work was funded at the University of Michigan through NSF:CPHOM (DMR 1120923). 

  \bibliography{paper14}

%merlin.mbs apsrev4-1.bst 2010-07-25 4.21a (PWD, AO, DPC) hacked
%Control: key (0)
%Control: author (8) initials jnrlst
%Control: editor formatted (1) identically to author
%Control: production of article title (-1) disabled
%Control: page (0) single
%Control: year (1) truncated
%Control: production of eprint (0) enabled
\begin{thebibliography}{52}%
\makeatletter
\providecommand \@ifxundefined [1]{%
 \@ifx{#1\undefined}
}%
\providecommand \@ifnum [1]{%
 \ifnum #1\expandafter \@firstoftwo
 \else \expandafter \@secondoftwo
 \fi
}%
\providecommand \@ifx [1]{%
 \ifx #1\expandafter \@firstoftwo
 \else \expandafter \@secondoftwo
 \fi
}%
\providecommand \natexlab [1]{#1}%
\providecommand \enquote  [1]{``#1''}%
\providecommand \bibnamefont  [1]{#1}%
\providecommand \bibfnamefont [1]{#1}%
\providecommand \citenamefont [1]{#1}%
\providecommand \href@noop [0]{\@secondoftwo}%
\providecommand \href [0]{\begingroup \@sanitize@url \@href}%
\providecommand \@href[1]{\@@startlink{#1}\@@href}%
\providecommand \@@href[1]{\endgroup#1\@@endlink}%
\providecommand \@sanitize@url [0]{\catcode `\\12\catcode `\$12\catcode
  `\&12\catcode `\#12\catcode `\^12\catcode `\_12\catcode `\%12\relax}%
\providecommand \@@startlink[1]{}%
\providecommand \@@endlink[0]{}%
\providecommand \url  [0]{\begingroup\@sanitize@url \@url }%
\providecommand \@url [1]{\endgroup\@href {#1}{\urlprefix }}%
\providecommand \urlprefix  [0]{URL }%
\providecommand \Eprint [0]{\href }%
\providecommand \doibase [0]{http://dx.doi.org/}%
\providecommand \selectlanguage [0]{\@gobble}%
\providecommand \bibinfo  [0]{\@secondoftwo}%
\providecommand \bibfield  [0]{\@secondoftwo}%
\providecommand \translation [1]{[#1]}%
\providecommand \BibitemOpen [0]{}%
\providecommand \bibitemStop [0]{}%
\providecommand \bibitemNoStop [0]{.\EOS\space}%
\providecommand \EOS [0]{\spacefactor3000\relax}%
\providecommand \BibitemShut  [1]{\csname bibitem#1\endcsname}%
\let\auto@bib@innerbib\@empty
%</preamble>
\bibitem [{\citenamefont {Holmes}\ \emph {et~al.}(2014)\citenamefont {Holmes},
  \citenamefont {Choi}, \citenamefont {Kako}, \citenamefont {Arita},\ and\
  \citenamefont {Arakawa}}]{doi:10.1021/nl404400d}%
  \BibitemOpen
  \bibfield  {author} {\bibinfo {author} {\bibfnamefont {M.~J.}\ \bibnamefont
  {Holmes}}, \bibinfo {author} {\bibfnamefont {K.}~\bibnamefont {Choi}},
  \bibinfo {author} {\bibfnamefont {S.}~\bibnamefont {Kako}}, \bibinfo {author}
  {\bibfnamefont {M.}~\bibnamefont {Arita}}, \ and\ \bibinfo {author}
  {\bibfnamefont {Y.}~\bibnamefont {Arakawa}},\ }\href {\doibase
  10.1021/nl404400d} {\bibfield  {journal} {\bibinfo  {journal} {Nano Lett.}\
  }\textbf {\bibinfo {volume} {14}},\ \bibinfo {pages} {982} (\bibinfo {year}
  {2014})}\BibitemShut {NoStop}%
\bibitem [{\citenamefont {Deshpande}\ \emph {et~al.}(2014)\citenamefont
  {Deshpande}, \citenamefont {Frost}, \citenamefont {Hazari},\ and\
  \citenamefont {Bhattacharya}}]{DeshpandeFrostHazariEtAl2014}%
  \BibitemOpen
  \bibfield  {author} {\bibinfo {author} {\bibfnamefont {S.}~\bibnamefont
  {Deshpande}}, \bibinfo {author} {\bibfnamefont {T.}~\bibnamefont {Frost}},
  \bibinfo {author} {\bibfnamefont {A.}~\bibnamefont {Hazari}}, \ and\ \bibinfo
  {author} {\bibfnamefont {P.}~\bibnamefont {Bhattacharya}},\ }\href {\doibase
  10.1063/1.4897640} {\bibfield  {journal} {\bibinfo  {journal} {Appl. Phys.
  Lett.}\ }\textbf {\bibinfo {volume} {105}},\ \bibinfo {eid} {141109}
  (\bibinfo {year} {2014}),\ 10.1063/1.4897640}\BibitemShut {NoStop}%
\bibitem [{\citenamefont {Bonadeo}\ \emph
  {et~al.}(1998{\natexlab{a}})\citenamefont {Bonadeo}, \citenamefont {Erland},
  \citenamefont {Gammon}, \citenamefont {Park}, \citenamefont {Katzer},\ and\
  \citenamefont {Steel}}]{BonadeoErlandGammonEtAl1998}%
  \BibitemOpen
  \bibfield  {author} {\bibinfo {author} {\bibfnamefont {N.~H.}\ \bibnamefont
  {Bonadeo}}, \bibinfo {author} {\bibfnamefont {J.}~\bibnamefont {Erland}},
  \bibinfo {author} {\bibfnamefont {D.}~\bibnamefont {Gammon}}, \bibinfo
  {author} {\bibfnamefont {D.}~\bibnamefont {Park}}, \bibinfo {author}
  {\bibfnamefont {D.~S.}\ \bibnamefont {Katzer}}, \ and\ \bibinfo {author}
  {\bibfnamefont {D.~G.}\ \bibnamefont {Steel}},\ }\href {\doibase
  10.1126/science.282.5393.1473} {\bibfield  {journal} {\bibinfo  {journal}
  {Science}\ }\textbf {\bibinfo {volume} {282}},\ \bibinfo {pages} {1473}
  (\bibinfo {year} {1998}{\natexlab{a}})}\BibitemShut {NoStop}%
\bibitem [{\citenamefont {Stievater}\ \emph {et~al.}(2001)\citenamefont
  {Stievater}, \citenamefont {Li}, \citenamefont {Steel}, \citenamefont
  {Gammon}, \citenamefont {Katzer}, \citenamefont {Park}, \citenamefont
  {Piermarocchi},\ and\ \citenamefont {Sham}}]{StievaterLiSteelEtAl2001}%
  \BibitemOpen
  \bibfield  {author} {\bibinfo {author} {\bibfnamefont {T.~H.}\ \bibnamefont
  {Stievater}}, \bibinfo {author} {\bibfnamefont {X.}~\bibnamefont {Li}},
  \bibinfo {author} {\bibfnamefont {D.~G.}\ \bibnamefont {Steel}}, \bibinfo
  {author} {\bibfnamefont {D.}~\bibnamefont {Gammon}}, \bibinfo {author}
  {\bibfnamefont {D.~S.}\ \bibnamefont {Katzer}}, \bibinfo {author}
  {\bibfnamefont {D.}~\bibnamefont {Park}}, \bibinfo {author} {\bibfnamefont
  {C.}~\bibnamefont {Piermarocchi}}, \ and\ \bibinfo {author} {\bibfnamefont
  {L.~J.}\ \bibnamefont {Sham}},\ }\href {\doibase
  10.1103/PhysRevLett.87.133603} {\bibfield  {journal} {\bibinfo  {journal}
  {Phys. Rev. Lett.}\ }\textbf {\bibinfo {volume} {87}},\ \bibinfo {pages}
  {133603} (\bibinfo {year} {2001})}\BibitemShut {NoStop}%
\bibitem [{\citenamefont {Kamada}\ \emph {et~al.}(2001)\citenamefont {Kamada},
  \citenamefont {Gotoh}, \citenamefont {Temmyo}, \citenamefont {Takagahara},\
  and\ \citenamefont {Ando}}]{KamadaGotohTemmyoEtAl2001}%
  \BibitemOpen
  \bibfield  {author} {\bibinfo {author} {\bibfnamefont {H.}~\bibnamefont
  {Kamada}}, \bibinfo {author} {\bibfnamefont {H.}~\bibnamefont {Gotoh}},
  \bibinfo {author} {\bibfnamefont {J.}~\bibnamefont {Temmyo}}, \bibinfo
  {author} {\bibfnamefont {T.}~\bibnamefont {Takagahara}}, \ and\ \bibinfo
  {author} {\bibfnamefont {H.}~\bibnamefont {Ando}},\ }\href {\doibase
  10.1103/PhysRevLett.87.246401} {\bibfield  {journal} {\bibinfo  {journal}
  {Phys. Rev. Lett.}\ }\textbf {\bibinfo {volume} {87}},\ \bibinfo {pages}
  {246401} (\bibinfo {year} {2001})}\BibitemShut {NoStop}%
\bibitem [{\citenamefont {Dutt}\ \emph {et~al.}(2005)\citenamefont {Dutt},
  \citenamefont {Cheng}, \citenamefont {Li}, \citenamefont {Xu}, \citenamefont
  {Li}, \citenamefont {Berman}, \citenamefont {Steel}, \citenamefont {Bracker},
  \citenamefont {Gammon}, \citenamefont {Economou}, \citenamefont {Liu},\ and\
  \citenamefont {Sham}}]{DuttChengLiEtAl2005}%
  \BibitemOpen
  \bibfield  {author} {\bibinfo {author} {\bibfnamefont {M.~V.~G.}\
  \bibnamefont {Dutt}}, \bibinfo {author} {\bibfnamefont {J.}~\bibnamefont
  {Cheng}}, \bibinfo {author} {\bibfnamefont {B.}~\bibnamefont {Li}}, \bibinfo
  {author} {\bibfnamefont {X.}~\bibnamefont {Xu}}, \bibinfo {author}
  {\bibfnamefont {X.}~\bibnamefont {Li}}, \bibinfo {author} {\bibfnamefont
  {P.~R.}\ \bibnamefont {Berman}}, \bibinfo {author} {\bibfnamefont {D.~G.}\
  \bibnamefont {Steel}}, \bibinfo {author} {\bibfnamefont {A.~S.}\ \bibnamefont
  {Bracker}}, \bibinfo {author} {\bibfnamefont {D.}~\bibnamefont {Gammon}},
  \bibinfo {author} {\bibfnamefont {S.~E.}\ \bibnamefont {Economou}}, \bibinfo
  {author} {\bibfnamefont {R.-B.}\ \bibnamefont {Liu}}, \ and\ \bibinfo
  {author} {\bibfnamefont {L.~J.}\ \bibnamefont {Sham}},\ }\href {\doibase
  10.1103/PhysRevLett.94.227403} {\bibfield  {journal} {\bibinfo  {journal}
  {Phys. Rev. Lett.}\ }\textbf {\bibinfo {volume} {94}},\ \bibinfo {pages}
  {227403} (\bibinfo {year} {2005})}\BibitemShut {NoStop}%
\bibitem [{\citenamefont {Santori}\ \emph {et~al.}(2006)\citenamefont
  {Santori}, \citenamefont {Tamarat}, \citenamefont {Neumann}, \citenamefont
  {Wrachtrup}, \citenamefont {Fattal}, \citenamefont {Beausoleil},
  \citenamefont {Rabeau}, \citenamefont {Olivero}, \citenamefont {Greentree},
  \citenamefont {Prawer}, \citenamefont {Jelezko},\ and\ \citenamefont
  {Hemmer}}]{SantoriTamaratNeumannEtAl2006}%
  \BibitemOpen
  \bibfield  {author} {\bibinfo {author} {\bibfnamefont {C.}~\bibnamefont
  {Santori}}, \bibinfo {author} {\bibfnamefont {P.}~\bibnamefont {Tamarat}},
  \bibinfo {author} {\bibfnamefont {P.}~\bibnamefont {Neumann}}, \bibinfo
  {author} {\bibfnamefont {J.}~\bibnamefont {Wrachtrup}}, \bibinfo {author}
  {\bibfnamefont {D.}~\bibnamefont {Fattal}}, \bibinfo {author} {\bibfnamefont
  {R.~G.}\ \bibnamefont {Beausoleil}}, \bibinfo {author} {\bibfnamefont
  {J.}~\bibnamefont {Rabeau}}, \bibinfo {author} {\bibfnamefont
  {P.}~\bibnamefont {Olivero}}, \bibinfo {author} {\bibfnamefont {A.~D.}\
  \bibnamefont {Greentree}}, \bibinfo {author} {\bibfnamefont {S.}~\bibnamefont
  {Prawer}}, \bibinfo {author} {\bibfnamefont {F.}~\bibnamefont {Jelezko}}, \
  and\ \bibinfo {author} {\bibfnamefont {P.}~\bibnamefont {Hemmer}},\ }\href
  {\doibase 10.1103/PhysRevLett.97.247401} {\bibfield  {journal} {\bibinfo
  {journal} {Phys. Rev. Lett.}\ }\textbf {\bibinfo {volume} {97}},\ \bibinfo
  {pages} {247401} (\bibinfo {year} {2006})}\BibitemShut {NoStop}%
\bibitem [{\citenamefont {Press}\ \emph {et~al.}(2008)\citenamefont {Press},
  \citenamefont {Ladd}, \citenamefont {Zhang},\ and\ \citenamefont
  {Yamamoto}}]{PressLaddZhangEtAl2008}%
  \BibitemOpen
  \bibfield  {author} {\bibinfo {author} {\bibfnamefont {D.}~\bibnamefont
  {Press}}, \bibinfo {author} {\bibfnamefont {T.~D.}\ \bibnamefont {Ladd}},
  \bibinfo {author} {\bibfnamefont {B.}~\bibnamefont {Zhang}}, \ and\ \bibinfo
  {author} {\bibfnamefont {Y.}~\bibnamefont {Yamamoto}},\ }\href {\doibase
  10.1038/nature07530} {\bibfield  {journal} {\bibinfo  {journal} {Nature}\
  }\textbf {\bibinfo {volume} {456}},\ \bibinfo {pages} {218} (\bibinfo {year}
  {2008})}\BibitemShut {NoStop}%
\bibitem [{\citenamefont {Holmes}\ \emph {et~al.}(2013)\citenamefont {Holmes},
  \citenamefont {Kako}, \citenamefont {Choi}, \citenamefont {Podemski},
  \citenamefont {Arita},\ and\ \citenamefont
  {Arakawa}}]{HolmesKakoChoiEtAl2013}%
  \BibitemOpen
  \bibfield  {author} {\bibinfo {author} {\bibfnamefont {M.}~\bibnamefont
  {Holmes}}, \bibinfo {author} {\bibfnamefont {S.}~\bibnamefont {Kako}},
  \bibinfo {author} {\bibfnamefont {K.}~\bibnamefont {Choi}}, \bibinfo {author}
  {\bibfnamefont {P.}~\bibnamefont {Podemski}}, \bibinfo {author}
  {\bibfnamefont {M.}~\bibnamefont {Arita}}, \ and\ \bibinfo {author}
  {\bibfnamefont {Y.}~\bibnamefont {Arakawa}},\ }\href {\doibase
  10.1103/PhysRevLett.111.057401} {\bibfield  {journal} {\bibinfo  {journal}
  {Phys. Rev. Lett.}\ }\textbf {\bibinfo {volume} {111}},\ \bibinfo {pages}
  {057401} (\bibinfo {year} {2013})}\BibitemShut {NoStop}%
\bibitem [{\citenamefont {Reid}\ \emph {et~al.}(2014)\citenamefont {Reid},
  \citenamefont {Kocher}, \citenamefont {Zhu}, \citenamefont {Oehler},
  \citenamefont {Emery}, \citenamefont {Chan}, \citenamefont {Oliver},\ and\
  \citenamefont {Taylor}}]{ReidKocherZhuEtAl2014}%
  \BibitemOpen
  \bibfield  {author} {\bibinfo {author} {\bibfnamefont {B.~P.~L.}\
  \bibnamefont {Reid}}, \bibinfo {author} {\bibfnamefont {C.}~\bibnamefont
  {Kocher}}, \bibinfo {author} {\bibfnamefont {T.}~\bibnamefont {Zhu}},
  \bibinfo {author} {\bibfnamefont {F.}~\bibnamefont {Oehler}}, \bibinfo
  {author} {\bibfnamefont {R.}~\bibnamefont {Emery}}, \bibinfo {author}
  {\bibfnamefont {C.~C.~S.}\ \bibnamefont {Chan}}, \bibinfo {author}
  {\bibfnamefont {R.~A.}\ \bibnamefont {Oliver}}, \ and\ \bibinfo {author}
  {\bibfnamefont {R.~A.}\ \bibnamefont {Taylor}},\ }\href {\doibase
  10.1063/1.4886961} {\bibfield  {journal} {\bibinfo  {journal} {Appl. Phys.
  Lett.}\ }\textbf {\bibinfo {volume} {104}},\ \bibinfo {eid} {263108}
  (\bibinfo {year} {2014}),\ 10.1063/1.4886961}\BibitemShut {NoStop}%
\bibitem [{\citenamefont {Hadfield}(2009)}]{Hadfield2009}%
  \BibitemOpen
  \bibfield  {author} {\bibinfo {author} {\bibfnamefont {R.~H.}\ \bibnamefont
  {Hadfield}},\ }\href {\doibase 10.1038/nphoton.2009.230} {\bibfield
  {journal} {\bibinfo  {journal} {Nat. Photon.}\ }\textbf {\bibinfo {volume}
  {3}},\ \bibinfo {pages} {696} (\bibinfo {year} {2009})}\BibitemShut {NoStop}%
\bibitem [{\citenamefont {Bonadeo}\ \emph
  {et~al.}(1998{\natexlab{b}})\citenamefont {Bonadeo}, \citenamefont {Chen},
  \citenamefont {Gammon}, \citenamefont {Katzer}, \citenamefont {Park},\ and\
  \citenamefont {Steel}}]{BonadeoChenGammonEtAl1998}%
  \BibitemOpen
  \bibfield  {author} {\bibinfo {author} {\bibfnamefont {N.~H.}\ \bibnamefont
  {Bonadeo}}, \bibinfo {author} {\bibfnamefont {G.}~\bibnamefont {Chen}},
  \bibinfo {author} {\bibfnamefont {D.}~\bibnamefont {Gammon}}, \bibinfo
  {author} {\bibfnamefont {D.~S.}\ \bibnamefont {Katzer}}, \bibinfo {author}
  {\bibfnamefont {D.}~\bibnamefont {Park}}, \ and\ \bibinfo {author}
  {\bibfnamefont {D.~G.}\ \bibnamefont {Steel}},\ }\href {\doibase
  10.1103/PhysRevLett.81.2759} {\bibfield  {journal} {\bibinfo  {journal}
  {Phys. Rev. Lett.}\ }\textbf {\bibinfo {volume} {81}},\ \bibinfo {pages}
  {2759} (\bibinfo {year} {1998}{\natexlab{b}})}\BibitemShut {NoStop}%
\bibitem [{\citenamefont {Wu}\ \emph {et~al.}(2004)\citenamefont {Wu},
  \citenamefont {Li}, \citenamefont {Steel}, \citenamefont {Gammon},\ and\
  \citenamefont {Sham}}]{WuLiSteelEtAl2004}%
  \BibitemOpen
  \bibfield  {author} {\bibinfo {author} {\bibfnamefont {Y.}~\bibnamefont
  {Wu}}, \bibinfo {author} {\bibfnamefont {X.}~\bibnamefont {Li}}, \bibinfo
  {author} {\bibfnamefont {D.}~\bibnamefont {Steel}}, \bibinfo {author}
  {\bibfnamefont {D.}~\bibnamefont {Gammon}}, \ and\ \bibinfo {author}
  {\bibfnamefont {L.}~\bibnamefont {Sham}},\ }\href {\doibase
  10.1016/j.physe.2004.06.023} {\bibfield  {journal} {\bibinfo  {journal}
  {Physica E}\ }\textbf {\bibinfo {volume} {25}},\ \bibinfo {pages} {242 }
  (\bibinfo {year} {2004})},\ \bibinfo {note} {proceedings of the 13th
  International Winterschool on New Developments in Solid State Physics -
  Low-Dimensional Systems}\BibitemShut {NoStop}%
\bibitem [{\citenamefont {Martin}\ \emph {et~al.}(1999)\citenamefont {Martin},
  \citenamefont {Middleton}, \citenamefont {Oâ€™Donnell},\ and\
  \citenamefont {Van~der Stricht}}]{MartinMiddletonOa€™DonnellEtAl1999}%
  \BibitemOpen
  \bibfield  {author} {\bibinfo {author} {\bibfnamefont {R.~W.}\ \bibnamefont
  {Martin}}, \bibinfo {author} {\bibfnamefont {P.~G.}\ \bibnamefont
  {Middleton}}, \bibinfo {author} {\bibfnamefont {K.~P.}\ \bibnamefont
  {Oâ€™Donnell}}, \ and\ \bibinfo {author} {\bibfnamefont
  {W.}~\bibnamefont {Van~der Stricht}},\ }\href {\doibase 10.1063/1.123275}
  {\bibfield  {journal} {\bibinfo  {journal} {Appl. Phys. Lett.}\ }\textbf
  {\bibinfo {volume} {74}},\ \bibinfo {pages} {263} (\bibinfo {year}
  {1999})}\BibitemShut {NoStop}%
\bibitem [{\citenamefont {Sch\"omig}\ \emph {et~al.}(2004)\citenamefont
  {Sch\"omig}, \citenamefont {Halm}, \citenamefont {Forchel}, \citenamefont
  {Bacher}, \citenamefont {Off},\ and\ \citenamefont
  {Scholz}}]{SchoemigHalmForchelEtAl2004}%
  \BibitemOpen
  \bibfield  {author} {\bibinfo {author} {\bibfnamefont {H.}~\bibnamefont
  {Sch\"omig}}, \bibinfo {author} {\bibfnamefont {S.}~\bibnamefont {Halm}},
  \bibinfo {author} {\bibfnamefont {A.}~\bibnamefont {Forchel}}, \bibinfo
  {author} {\bibfnamefont {G.}~\bibnamefont {Bacher}}, \bibinfo {author}
  {\bibfnamefont {J.}~\bibnamefont {Off}}, \ and\ \bibinfo {author}
  {\bibfnamefont {F.}~\bibnamefont {Scholz}},\ }\href {\doibase
  10.1103/PhysRevLett.92.106802} {\bibfield  {journal} {\bibinfo  {journal}
  {Phys. Rev. Lett.}\ }\textbf {\bibinfo {volume} {92}},\ \bibinfo {pages}
  {106802} (\bibinfo {year} {2004})}\BibitemShut {NoStop}%
\bibitem [{\citenamefont {Woo}\ \emph {et~al.}(2015)\citenamefont {Woo},
  \citenamefont {Bugnet}, \citenamefont {Nguyen}, \citenamefont {Mi},\ and\
  \citenamefont {Botton}}]{WooBugnetNguyenEtAl2015}%
  \BibitemOpen
  \bibfield  {author} {\bibinfo {author} {\bibfnamefont {S.~Y.}\ \bibnamefont
  {Woo}}, \bibinfo {author} {\bibfnamefont {M.}~\bibnamefont {Bugnet}},
  \bibinfo {author} {\bibfnamefont {H.~P.~T.}\ \bibnamefont {Nguyen}}, \bibinfo
  {author} {\bibfnamefont {Z.}~\bibnamefont {Mi}}, \ and\ \bibinfo {author}
  {\bibfnamefont {G.~A.}\ \bibnamefont {Botton}},\ }\href {\doibase
  10.1021/acs.nanolett.5b01628} {\bibfield  {journal} {\bibinfo  {journal}
  {Nano Lett.}\ }\textbf {\bibinfo {volume} {15}},\ \bibinfo {pages} {6413}
  (\bibinfo {year} {2015})}\BibitemShut {NoStop}%
\bibitem [{\citenamefont {Li}\ \emph {et~al.}(2010)\citenamefont {Li},
  \citenamefont {Fischer}, \citenamefont {Wei}, \citenamefont {Ponce},
  \citenamefont {Detchprohm},\ and\ \citenamefont
  {Wetzel}}]{LiFischerWeiEtAl2010}%
  \BibitemOpen
  \bibfield  {author} {\bibinfo {author} {\bibfnamefont {T.}~\bibnamefont
  {Li}}, \bibinfo {author} {\bibfnamefont {A.~M.}\ \bibnamefont {Fischer}},
  \bibinfo {author} {\bibfnamefont {Q.~Y.}\ \bibnamefont {Wei}}, \bibinfo
  {author} {\bibfnamefont {F.~A.}\ \bibnamefont {Ponce}}, \bibinfo {author}
  {\bibfnamefont {T.}~\bibnamefont {Detchprohm}}, \ and\ \bibinfo {author}
  {\bibfnamefont {C.}~\bibnamefont {Wetzel}},\ }\href {\doibase
  10.1063/1.3293298} {\bibfield  {journal} {\bibinfo  {journal} {Appl. Phys.
  Lett.}\ }\textbf {\bibinfo {volume} {96}},\ \bibinfo {eid} {031906} (\bibinfo
  {year} {2010}),\ 10.1063/1.3293298}\BibitemShut {NoStop}%
\bibitem [{\citenamefont {Schulz}\ \emph {et~al.}(2015)\citenamefont {Schulz},
  \citenamefont {Caro}, \citenamefont {Coughlan},\ and\ \citenamefont
  {O'Reilly}}]{SchulzCaroCoughlanEtAl2015}%
  \BibitemOpen
  \bibfield  {author} {\bibinfo {author} {\bibfnamefont {S.}~\bibnamefont
  {Schulz}}, \bibinfo {author} {\bibfnamefont {M.~A.}\ \bibnamefont {Caro}},
  \bibinfo {author} {\bibfnamefont {C.}~\bibnamefont {Coughlan}}, \ and\
  \bibinfo {author} {\bibfnamefont {E.~P.}\ \bibnamefont {O'Reilly}},\ }\href
  {\doibase 10.1103/PhysRevB.91.035439} {\bibfield  {journal} {\bibinfo
  {journal} {Phys. Rev. B}\ }\textbf {\bibinfo {volume} {91}},\ \bibinfo
  {pages} {035439} (\bibinfo {year} {2015})}\BibitemShut {NoStop}%
\bibitem [{\citenamefont {Chichibu}\ \emph {et~al.}(2006)\citenamefont
  {Chichibu}, \citenamefont {Uedono}, \citenamefont {Onuma}, \citenamefont
  {Haskell}, \citenamefont {Chakraborty}, \citenamefont {Koyama}, \citenamefont
  {Fini}, \citenamefont {Keller}, \citenamefont {DenBaars}, \citenamefont
  {Speck}, \citenamefont {Mishra}, \citenamefont {Nakamura}, \citenamefont
  {Yamaguchi}, \citenamefont {Kamiyama}, \citenamefont {Amano}, \citenamefont
  {Akasaki}, \citenamefont {Han},\ and\ \citenamefont
  {Sota}}]{ChichibuUedonoOnumaEtAl2006}%
  \BibitemOpen
  \bibfield  {author} {\bibinfo {author} {\bibfnamefont {S.~F.}\ \bibnamefont
  {Chichibu}}, \bibinfo {author} {\bibfnamefont {A.}~\bibnamefont {Uedono}},
  \bibinfo {author} {\bibfnamefont {T.}~\bibnamefont {Onuma}}, \bibinfo
  {author} {\bibfnamefont {B.~A.}\ \bibnamefont {Haskell}}, \bibinfo {author}
  {\bibfnamefont {A.}~\bibnamefont {Chakraborty}}, \bibinfo {author}
  {\bibfnamefont {T.}~\bibnamefont {Koyama}}, \bibinfo {author} {\bibfnamefont
  {P.~T.}\ \bibnamefont {Fini}}, \bibinfo {author} {\bibfnamefont
  {S.}~\bibnamefont {Keller}}, \bibinfo {author} {\bibfnamefont {S.~P.}\
  \bibnamefont {DenBaars}}, \bibinfo {author} {\bibfnamefont {J.~S.}\
  \bibnamefont {Speck}}, \bibinfo {author} {\bibfnamefont {U.~K.}\ \bibnamefont
  {Mishra}}, \bibinfo {author} {\bibfnamefont {S.}~\bibnamefont {Nakamura}},
  \bibinfo {author} {\bibfnamefont {S.}~\bibnamefont {Yamaguchi}}, \bibinfo
  {author} {\bibfnamefont {S.}~\bibnamefont {Kamiyama}}, \bibinfo {author}
  {\bibfnamefont {H.}~\bibnamefont {Amano}}, \bibinfo {author} {\bibfnamefont
  {I.}~\bibnamefont {Akasaki}}, \bibinfo {author} {\bibfnamefont
  {J.}~\bibnamefont {Han}}, \ and\ \bibinfo {author} {\bibfnamefont
  {T.}~\bibnamefont {Sota}},\ }\href {\doibase 10.1038/nmat1726} {\bibfield
  {journal} {\bibinfo  {journal} {Nat. Mater.}\ }\textbf {\bibinfo {volume}
  {5}},\ \bibinfo {pages} {810} (\bibinfo {year} {2006})}\BibitemShut {NoStop}%
\bibitem [{\citenamefont {Stryland}\ \emph {et~al.}(1988)\citenamefont
  {Stryland}, \citenamefont {Wu}, \citenamefont {Hagan}, \citenamefont
  {Soileau},\ and\ \citenamefont {Mansour}}]{StrylandWuHaganEtAl1988}%
  \BibitemOpen
  \bibfield  {author} {\bibinfo {author} {\bibfnamefont {E.~W.~V.}\
  \bibnamefont {Stryland}}, \bibinfo {author} {\bibfnamefont {Y.~Y.}\
  \bibnamefont {Wu}}, \bibinfo {author} {\bibfnamefont {D.~J.}\ \bibnamefont
  {Hagan}}, \bibinfo {author} {\bibfnamefont {M.~J.}\ \bibnamefont {Soileau}},
  \ and\ \bibinfo {author} {\bibfnamefont {K.}~\bibnamefont {Mansour}},\ }\href
  {\doibase 10.1364/JOSAB.5.001980} {\bibfield  {journal} {\bibinfo  {journal}
  {J. Opt. Soc. Am. B}\ }\textbf {\bibinfo {volume} {5}},\ \bibinfo {pages}
  {1980} (\bibinfo {year} {1988})}\BibitemShut {NoStop}%
\bibitem [{\citenamefont {Kundys}\ \emph {et~al.}(2006)\citenamefont {Kundys},
  \citenamefont {Wells}, \citenamefont {Andreev}, \citenamefont {Hashemizadeh},
  \citenamefont {Wang}, \citenamefont {Parbrook}, \citenamefont {Fox},
  \citenamefont {Mowbray},\ and\ \citenamefont
  {Skolnick}}]{KundysWellsAndreevEtAl2006}%
  \BibitemOpen
  \bibfield  {author} {\bibinfo {author} {\bibfnamefont {D.~O.}\ \bibnamefont
  {Kundys}}, \bibinfo {author} {\bibfnamefont {J.-P.~R.}\ \bibnamefont
  {Wells}}, \bibinfo {author} {\bibfnamefont {A.~D.}\ \bibnamefont {Andreev}},
  \bibinfo {author} {\bibfnamefont {S.~A.}\ \bibnamefont {Hashemizadeh}},
  \bibinfo {author} {\bibfnamefont {T.}~\bibnamefont {Wang}}, \bibinfo {author}
  {\bibfnamefont {P.~J.}\ \bibnamefont {Parbrook}}, \bibinfo {author}
  {\bibfnamefont {A.~M.}\ \bibnamefont {Fox}}, \bibinfo {author} {\bibfnamefont
  {D.~J.}\ \bibnamefont {Mowbray}}, \ and\ \bibinfo {author} {\bibfnamefont
  {M.~S.}\ \bibnamefont {Skolnick}},\ }\href {\doibase
  10.1103/PhysRevB.73.165309} {\bibfield  {journal} {\bibinfo  {journal} {Phys.
  Rev. B}\ }\textbf {\bibinfo {volume} {73}},\ \bibinfo {pages} {165309}
  (\bibinfo {year} {2006})}\BibitemShut {NoStop}%
\bibitem [{\citenamefont {Guo}\ \emph {et~al.}(2010)\citenamefont {Guo},
  \citenamefont {Zhang}, \citenamefont {Banerjee},\ and\ \citenamefont
  {Bhattacharya}}]{GuoZhangBanerjeeEtAl2010}%
  \BibitemOpen
  \bibfield  {author} {\bibinfo {author} {\bibfnamefont {W.}~\bibnamefont
  {Guo}}, \bibinfo {author} {\bibfnamefont {M.}~\bibnamefont {Zhang}}, \bibinfo
  {author} {\bibfnamefont {A.}~\bibnamefont {Banerjee}}, \ and\ \bibinfo
  {author} {\bibfnamefont {P.}~\bibnamefont {Bhattacharya}},\ }\href {\doibase
  10.1021/nl101027x} {\bibfield  {journal} {\bibinfo  {journal} {Nano Lett.}\
  }\textbf {\bibinfo {volume} {10}},\ \bibinfo {pages} {3355} (\bibinfo {year}
  {2010})}\BibitemShut {NoStop}%
\bibitem [{\citenamefont {Jahangir}\ \emph {et~al.}(2013)\citenamefont
  {Jahangir}, \citenamefont {Mandl}, \citenamefont {Strassburg},\ and\
  \citenamefont {Bhattacharya}}]{JahangirMandlStrassburgEtAl2013}%
  \BibitemOpen
  \bibfield  {author} {\bibinfo {author} {\bibfnamefont {S.}~\bibnamefont
  {Jahangir}}, \bibinfo {author} {\bibfnamefont {M.}~\bibnamefont {Mandl}},
  \bibinfo {author} {\bibfnamefont {M.}~\bibnamefont {Strassburg}}, \ and\
  \bibinfo {author} {\bibfnamefont {P.}~\bibnamefont {Bhattacharya}},\ }\href
  {\doibase 10.1063/1.4793300} {\bibfield  {journal} {\bibinfo  {journal}
  {Appl. Phys. Lett.}\ }\textbf {\bibinfo {volume} {102}},\ \bibinfo {eid}
  {071101} (\bibinfo {year} {2013}),\ 10.1063/1.4793300}\BibitemShut {NoStop}%
\bibitem [{\citenamefont {Deshpande}\ \emph {et~al.}(2015)\citenamefont
  {Deshpande}, \citenamefont {Frost}, \citenamefont {Yan}, \citenamefont
  {Jahangir}, \citenamefont {Hazari}, \citenamefont {Liu}, \citenamefont
  {Mirecki-Millunchick}, \citenamefont {Mi},\ and\ \citenamefont
  {Bhattacharya}}]{DeshpandeFrostYanEtAl2015}%
  \BibitemOpen
  \bibfield  {author} {\bibinfo {author} {\bibfnamefont {S.}~\bibnamefont
  {Deshpande}}, \bibinfo {author} {\bibfnamefont {T.}~\bibnamefont {Frost}},
  \bibinfo {author} {\bibfnamefont {L.}~\bibnamefont {Yan}}, \bibinfo {author}
  {\bibfnamefont {S.}~\bibnamefont {Jahangir}}, \bibinfo {author}
  {\bibfnamefont {A.}~\bibnamefont {Hazari}}, \bibinfo {author} {\bibfnamefont
  {X.}~\bibnamefont {Liu}}, \bibinfo {author} {\bibfnamefont {J.}~\bibnamefont
  {Mirecki-Millunchick}}, \bibinfo {author} {\bibfnamefont {Z.}~\bibnamefont
  {Mi}}, \ and\ \bibinfo {author} {\bibfnamefont {P.}~\bibnamefont
  {Bhattacharya}},\ }\href {\doibase 10.1021/nl5041989} {\bibfield  {journal}
  {\bibinfo  {journal} {Nano Lett.}\ }\textbf {\bibinfo {volume} {15}},\
  \bibinfo {pages} {1647} (\bibinfo {year} {2015})}\BibitemShut {NoStop}%
\bibitem [{\citenamefont {Tourbot}\ \emph {et~al.}(2011)\citenamefont
  {Tourbot}, \citenamefont {Bougerol}, \citenamefont {Grenier}, \citenamefont
  {Hertog}, \citenamefont {Sam-Giao}, \citenamefont {Cooper}, \citenamefont
  {Gilet}, \citenamefont {Gayral},\ and\ \citenamefont
  {Daudin}}]{TourbotBougerolGrenierEtAl2011}%
  \BibitemOpen
  \bibfield  {author} {\bibinfo {author} {\bibfnamefont {G.}~\bibnamefont
  {Tourbot}}, \bibinfo {author} {\bibfnamefont {C.}~\bibnamefont {Bougerol}},
  \bibinfo {author} {\bibfnamefont {A.}~\bibnamefont {Grenier}}, \bibinfo
  {author} {\bibfnamefont {M.~D.}\ \bibnamefont {Hertog}}, \bibinfo {author}
  {\bibfnamefont {D.}~\bibnamefont {Sam-Giao}}, \bibinfo {author}
  {\bibfnamefont {D.}~\bibnamefont {Cooper}}, \bibinfo {author} {\bibfnamefont
  {P.}~\bibnamefont {Gilet}}, \bibinfo {author} {\bibfnamefont
  {B.}~\bibnamefont {Gayral}}, \ and\ \bibinfo {author} {\bibfnamefont
  {B.}~\bibnamefont {Daudin}},\ }\href@noop {} {\bibfield  {journal} {\bibinfo
  {journal} {Nanotech.}\ }\textbf {\bibinfo {volume} {22}},\ \bibinfo {pages}
  {075601} (\bibinfo {year} {2011})}\BibitemShut {NoStop}%
\bibitem [{\citenamefont {Tourbot}\ \emph {et~al.}(2012)\citenamefont
  {Tourbot}, \citenamefont {Bougerol}, \citenamefont {Glas}, \citenamefont
  {Zagonel}, \citenamefont {Mahfoud}, \citenamefont {Meuret}, \citenamefont
  {Gilet}, \citenamefont {Kociak}, \citenamefont {Gayral},\ and\ \citenamefont
  {Daudin}}]{TourbotBougerolGlasEtAl2012}%
  \BibitemOpen
  \bibfield  {author} {\bibinfo {author} {\bibfnamefont {G.}~\bibnamefont
  {Tourbot}}, \bibinfo {author} {\bibfnamefont {C.}~\bibnamefont {Bougerol}},
  \bibinfo {author} {\bibfnamefont {F.}~\bibnamefont {Glas}}, \bibinfo {author}
  {\bibfnamefont {L.~F.}\ \bibnamefont {Zagonel}}, \bibinfo {author}
  {\bibfnamefont {Z.}~\bibnamefont {Mahfoud}}, \bibinfo {author} {\bibfnamefont
  {S.}~\bibnamefont {Meuret}}, \bibinfo {author} {\bibfnamefont
  {P.}~\bibnamefont {Gilet}}, \bibinfo {author} {\bibfnamefont
  {M.}~\bibnamefont {Kociak}}, \bibinfo {author} {\bibfnamefont
  {B.}~\bibnamefont {Gayral}}, \ and\ \bibinfo {author} {\bibfnamefont
  {B.}~\bibnamefont {Daudin}},\ }\href@noop {} {\bibfield  {journal} {\bibinfo
  {journal} {Nanotech.}\ }\textbf {\bibinfo {volume} {23}},\ \bibinfo {pages}
  {135703} (\bibinfo {year} {2012})}\BibitemShut {NoStop}%
\bibitem [{\citenamefont {Chang}\ \emph {et~al.}(2010)\citenamefont {Chang},
  \citenamefont {Wang}, \citenamefont {Li},\ and\ \citenamefont
  {Mi}}]{ChangWangLiEtAl2010}%
  \BibitemOpen
  \bibfield  {author} {\bibinfo {author} {\bibfnamefont {Y.-L.}\ \bibnamefont
  {Chang}}, \bibinfo {author} {\bibfnamefont {J.~L.}\ \bibnamefont {Wang}},
  \bibinfo {author} {\bibfnamefont {F.}~\bibnamefont {Li}}, \ and\ \bibinfo
  {author} {\bibfnamefont {Z.}~\bibnamefont {Mi}},\ }\href {\doibase
  10.1063/1.3284660} {\bibfield  {journal} {\bibinfo  {journal} {Appl. Phys.
  Lett.}\ }\textbf {\bibinfo {volume} {96}},\ \bibinfo {eid} {013106} (\bibinfo
  {year} {2010}),\ 10.1063/1.3284660}\BibitemShut {NoStop}%
\bibitem [{Sup()}]{Supp.Mat.}%
  \BibitemOpen
  \href@noop {} {\enquote {\bibinfo {title} {See supplemental material at [],
  which includes refs. [39-41,42,43,45] and a detailed theoretical
  consideration of some concepts in the main text as well as supplementary
  figures and experimental techniques.}}\ }\BibitemShut {NoStop}%
\bibitem [{\citenamefont {Yan}\ \emph {et~al.}(2015)\citenamefont {Yan},
  \citenamefont {Jahangir}, \citenamefont {Wight}, \citenamefont {Nikoobakht},
  \citenamefont {Bhattacharya},\ and\ \citenamefont
  {Millunchick}}]{YanJahangirWightEtAl2015}%
  \BibitemOpen
  \bibfield  {author} {\bibinfo {author} {\bibfnamefont {L.}~\bibnamefont
  {Yan}}, \bibinfo {author} {\bibfnamefont {S.}~\bibnamefont {Jahangir}},
  \bibinfo {author} {\bibfnamefont {S.~A.}\ \bibnamefont {Wight}}, \bibinfo
  {author} {\bibfnamefont {B.}~\bibnamefont {Nikoobakht}}, \bibinfo {author}
  {\bibfnamefont {P.}~\bibnamefont {Bhattacharya}}, \ and\ \bibinfo {author}
  {\bibfnamefont {J.~M.}\ \bibnamefont {Millunchick}},\ }\href {\doibase
  10.1021/nl503826k} {\bibfield  {journal} {\bibinfo  {journal} {Nano Lett.}\
  }\textbf {\bibinfo {volume} {15}},\ \bibinfo {pages} {1535} (\bibinfo {year}
  {2015})}\BibitemShut {NoStop}%
\bibitem [{\citenamefont {Sekiguchi}\ \emph {et~al.}(2010)\citenamefont
  {Sekiguchi}, \citenamefont {Kishino},\ and\ \citenamefont
  {Kikuchi}}]{SekiguchiKishinoKikuchi2010}%
  \BibitemOpen
  \bibfield  {author} {\bibinfo {author} {\bibfnamefont {H.}~\bibnamefont
  {Sekiguchi}}, \bibinfo {author} {\bibfnamefont {K.}~\bibnamefont {Kishino}},
  \ and\ \bibinfo {author} {\bibfnamefont {A.}~\bibnamefont {Kikuchi}},\ }\href
  {\doibase 10.1063/1.3443734} {\bibfield  {journal} {\bibinfo  {journal}
  {Appl. Phys. Lett.}\ }\textbf {\bibinfo {volume} {96}},\ \bibinfo {eid}
  {231104} (\bibinfo {year} {2010}),\ 10.1063/1.3443734}\BibitemShut {NoStop}%
\bibitem [{\citenamefont {Zhang}\ \emph {et~al.}(2013)\citenamefont {Zhang},
  \citenamefont {Teng}, \citenamefont {Hill}, \citenamefont {Lee},
  \citenamefont {Ku},\ and\ \citenamefont {Deng}}]{ZhangTengHillEtAl2013}%
  \BibitemOpen
  \bibfield  {author} {\bibinfo {author} {\bibfnamefont {L.}~\bibnamefont
  {Zhang}}, \bibinfo {author} {\bibfnamefont {C.-H.}\ \bibnamefont {Teng}},
  \bibinfo {author} {\bibfnamefont {T.~A.}\ \bibnamefont {Hill}}, \bibinfo
  {author} {\bibfnamefont {L.-K.}\ \bibnamefont {Lee}}, \bibinfo {author}
  {\bibfnamefont {P.-C.}\ \bibnamefont {Ku}}, \ and\ \bibinfo {author}
  {\bibfnamefont {H.}~\bibnamefont {Deng}},\ }\href {\doibase
  10.1063/1.4830000} {\bibfield  {journal} {\bibinfo  {journal} {Appl. Phys.
  Lett.}\ }\textbf {\bibinfo {volume} {103}},\ \bibinfo {eid} {192114}
  (\bibinfo {year} {2013}),\ 10.1063/1.4830000}\BibitemShut {NoStop}%
\bibitem [{\citenamefont {Sacconi}\ \emph {et~al.}(2012)\citenamefont
  {Sacconi}, \citenamefont {Auf~der Maur},\ and\ \citenamefont
  {Di~Carlo}}]{SacconiAufderMaurDiCarlo2012}%
  \BibitemOpen
  \bibfield  {author} {\bibinfo {author} {\bibfnamefont {F.}~\bibnamefont
  {Sacconi}}, \bibinfo {author} {\bibfnamefont {M.}~\bibnamefont {Auf~der
  Maur}}, \ and\ \bibinfo {author} {\bibfnamefont {A.}~\bibnamefont
  {Di~Carlo}},\ }\href {\doibase 10.1109/TED.2012.2210897} {\bibfield
  {journal} {\bibinfo  {journal} {IEEE Trans. Elect. Dev.}\ }\textbf {\bibinfo
  {volume} {59}},\ \bibinfo {pages} {2979} (\bibinfo {year}
  {2012})}\BibitemShut {NoStop}%
\bibitem [{\citenamefont {Zhang}\ \emph {et~al.}(2014)\citenamefont {Zhang},
  \citenamefont {Hill}, \citenamefont {Teng}, \citenamefont {Demory},
  \citenamefont {Ku},\ and\ \citenamefont {Deng}}]{ZhangHillTengEtAl2014}%
  \BibitemOpen
  \bibfield  {author} {\bibinfo {author} {\bibfnamefont {L.}~\bibnamefont
  {Zhang}}, \bibinfo {author} {\bibfnamefont {T.~A.}\ \bibnamefont {Hill}},
  \bibinfo {author} {\bibfnamefont {C.-H.}\ \bibnamefont {Teng}}, \bibinfo
  {author} {\bibfnamefont {B.}~\bibnamefont {Demory}}, \bibinfo {author}
  {\bibfnamefont {P.-C.}\ \bibnamefont {Ku}}, \ and\ \bibinfo {author}
  {\bibfnamefont {H.}~\bibnamefont {Deng}},\ }\href {\doibase
  10.1103/PhysRevB.90.245311} {\bibfield  {journal} {\bibinfo  {journal} {Phys.
  Rev. B}\ }\textbf {\bibinfo {volume} {90}},\ \bibinfo {pages} {245311}
  (\bibinfo {year} {2014})}\BibitemShut {NoStop}%
\bibitem [{\citenamefont {Pereira}\ \emph {et~al.}(2001)\citenamefont
  {Pereira}, \citenamefont {Correia}, \citenamefont {Pereira}, \citenamefont
  {O'Donnell}, \citenamefont {Trager-Cowan}, \citenamefont {Sweeney},\ and\
  \citenamefont {Alves}}]{PereiraCorreiaPereiraEtAl2001}%
  \BibitemOpen
  \bibfield  {author} {\bibinfo {author} {\bibfnamefont {S.}~\bibnamefont
  {Pereira}}, \bibinfo {author} {\bibfnamefont {M.~R.}\ \bibnamefont
  {Correia}}, \bibinfo {author} {\bibfnamefont {E.}~\bibnamefont {Pereira}},
  \bibinfo {author} {\bibfnamefont {K.~P.}\ \bibnamefont {O'Donnell}}, \bibinfo
  {author} {\bibfnamefont {C.}~\bibnamefont {Trager-Cowan}}, \bibinfo {author}
  {\bibfnamefont {F.}~\bibnamefont {Sweeney}}, \ and\ \bibinfo {author}
  {\bibfnamefont {E.}~\bibnamefont {Alves}},\ }\href {\doibase
  10.1103/PhysRevB.64.205311} {\bibfield  {journal} {\bibinfo  {journal} {Phys.
  Rev. B}\ }\textbf {\bibinfo {volume} {64}},\ \bibinfo {pages} {205311}
  (\bibinfo {year} {2001})}\BibitemShut {NoStop}%
\bibitem [{\citenamefont {Medhekar}\ \emph {et~al.}(2008)\citenamefont
  {Medhekar}, \citenamefont {Hegadekatte},\ and\ \citenamefont
  {Shenoy}}]{MedhekarHegadekatteShenoy2008}%
  \BibitemOpen
  \bibfield  {author} {\bibinfo {author} {\bibfnamefont {N.~V.}\ \bibnamefont
  {Medhekar}}, \bibinfo {author} {\bibfnamefont {V.}~\bibnamefont
  {Hegadekatte}}, \ and\ \bibinfo {author} {\bibfnamefont {V.~B.}\ \bibnamefont
  {Shenoy}},\ }\href {\doibase 10.1103/PhysRevLett.100.106104} {\bibfield
  {journal} {\bibinfo  {journal} {Phys. Rev. Lett.}\ }\textbf {\bibinfo
  {volume} {100}},\ \bibinfo {pages} {106104} (\bibinfo {year}
  {2008})}\BibitemShut {NoStop}%
\bibitem [{\citenamefont {Deshpande}\ \emph {et~al.}(2013)\citenamefont
  {Deshpande}, \citenamefont {Heo}, \citenamefont {Das},\ and\ \citenamefont
  {Bhattacharya}}]{DeshpandeHeoDasEtAl2013}%
  \BibitemOpen
  \bibfield  {author} {\bibinfo {author} {\bibfnamefont {S.}~\bibnamefont
  {Deshpande}}, \bibinfo {author} {\bibfnamefont {J.}~\bibnamefont {Heo}},
  \bibinfo {author} {\bibfnamefont {A.}~\bibnamefont {Das}}, \ and\ \bibinfo
  {author} {\bibfnamefont {P.}~\bibnamefont {Bhattacharya}},\ }\href {\doibase
  10.1038/ncomms2691} {\bibfield  {journal} {\bibinfo  {journal} {Nat.
  Commun.}\ }\textbf {\bibinfo {volume} {4}},\ \bibinfo {pages} {1675}
  (\bibinfo {year} {2013})}\BibitemShut {NoStop}%
\bibitem [{\citenamefont {Varshni}(1967)}]{Varshni1967}%
  \BibitemOpen
  \bibfield  {author} {\bibinfo {author} {\bibfnamefont {Y.}~\bibnamefont
  {Varshni}},\ }\href {\doibase 10.1016/0031-8914(67)90062-6} {\bibfield
  {journal} {\bibinfo  {journal} {Physica}\ }\textbf {\bibinfo {volume} {34}},\
  \bibinfo {pages} {149 } (\bibinfo {year} {1967})}\BibitemShut {NoStop}%
\bibitem [{\citenamefont {Vurgaftman}\ \emph {et~al.}(2001)\citenamefont
  {Vurgaftman}, \citenamefont {Meyer},\ and\ \citenamefont
  {Ram-Mohan}}]{VurgaftmanMeyerRam-Mohan2001}%
  \BibitemOpen
  \bibfield  {author} {\bibinfo {author} {\bibfnamefont {I.}~\bibnamefont
  {Vurgaftman}}, \bibinfo {author} {\bibfnamefont {J.~R.}\ \bibnamefont
  {Meyer}}, \ and\ \bibinfo {author} {\bibfnamefont {L.~R.}\ \bibnamefont
  {Ram-Mohan}},\ }\href {\doibase 10.1063/1.1368156} {\bibfield  {journal}
  {\bibinfo  {journal} {J. Appl. Phys.}\ }\textbf {\bibinfo {volume} {89}},\
  \bibinfo {pages} {5815} (\bibinfo {year} {2001})}\BibitemShut {NoStop}%
\bibitem [{\citenamefont {Jameson}\ \emph {et~al.}(1984)\citenamefont
  {Jameson}, \citenamefont {Gratton},\ and\ \citenamefont
  {Hall}}]{JamesonGrattonHall1984}%
  \BibitemOpen
  \bibfield  {author} {\bibinfo {author} {\bibfnamefont {D.~M.}\ \bibnamefont
  {Jameson}}, \bibinfo {author} {\bibfnamefont {E.}~\bibnamefont {Gratton}}, \
  and\ \bibinfo {author} {\bibfnamefont {R.~D.}\ \bibnamefont {Hall}},\ }\href
  {\doibase 10.1080/05704928408081716} {\bibfield  {journal} {\bibinfo
  {journal} {Appl. Spec. Rev.}\ }\textbf {\bibinfo {volume} {20}},\ \bibinfo
  {pages} {55} (\bibinfo {year} {1984})}\BibitemShut {NoStop}%
\bibitem [{\citenamefont {Lakowicz}(1999)}]{PrinFlSpec}%
  \BibitemOpen
  \bibfield  {author} {\bibinfo {author} {\bibfnamefont {J.~R.}\ \bibnamefont
  {Lakowicz}},\ }\href {\doibase 10.1007/978-0-387-46312-4} {\emph {\bibinfo
  {title} {Principles of Fluorescence Spectroscopy}}},\ \bibinfo {edition}
  {3rd}\ ed.\ (\bibinfo  {publisher} {Kluwer Academic, New York, NY, USA},\
  \bibinfo {year} {1999})\BibitemShut {NoStop}%
\bibitem [{\citenamefont {Umesh K.~Mishra}(2008)}]{SemDevDes}%
  \BibitemOpen
  \bibfield  {author} {\bibinfo {author} {\bibfnamefont {J.~S.}\ \bibnamefont
  {Umesh K.~Mishra}},\ }\href {\doibase 10.1007/978-1-4020-6481-4} {\emph
  {\bibinfo {title} {Semiconductor Device Physics and Design}}},\ \bibinfo
  {edition} {1st}\ ed.\ (\bibinfo  {publisher} {Springer, the Netherlands},\
  \bibinfo {year} {2008})\BibitemShut {NoStop}%
\bibitem [{\citenamefont {Steel}\ and\ \citenamefont
  {Rand}(1985)}]{SteelRand1985}%
  \BibitemOpen
  \bibfield  {author} {\bibinfo {author} {\bibfnamefont {D.~G.}\ \bibnamefont
  {Steel}}\ and\ \bibinfo {author} {\bibfnamefont {S.~C.}\ \bibnamefont
  {Rand}},\ }\href {\doibase 10.1103/PhysRevLett.55.2285} {\bibfield  {journal}
  {\bibinfo  {journal} {Phys. Rev. Lett.}\ }\textbf {\bibinfo {volume} {55}},\
  \bibinfo {pages} {2285} (\bibinfo {year} {1985})}\BibitemShut {NoStop}%
\bibitem [{\citenamefont {Lamb}(1964)}]{Lamb1964}%
  \BibitemOpen
  \bibfield  {author} {\bibinfo {author} {\bibfnamefont {W.~E.}\ \bibnamefont
  {Lamb}},\ }\href {\doibase 10.1103/PhysRev.134.A1429} {\bibfield  {journal}
  {\bibinfo  {journal} {Phys. Rev.}\ }\textbf {\bibinfo {volume} {134}},\
  \bibinfo {pages} {A1429} (\bibinfo {year} {1964})}\BibitemShut {NoStop}%
\bibitem [{\citenamefont {Shen}(1984)}]{ShenNLOpt}%
  \BibitemOpen
  \bibfield  {author} {\bibinfo {author} {\bibfnamefont {Y.}~\bibnamefont
  {Shen}},\ }\href@noop {} {\emph {\bibinfo {title} {Principles of non-linear
  optics}}}\ (\bibinfo  {publisher} {Wiley-Interscience, New York, NY, USA},\
  \bibinfo {year} {1984})\BibitemShut {NoStop}%
\bibitem [{\citenamefont {Berman}\ and\ \citenamefont
  {Malinovsky}(2011)}]{Berman}%
  \BibitemOpen
  \bibfield  {author} {\bibinfo {author} {\bibfnamefont {P.~R.}\ \bibnamefont
  {Berman}}\ and\ \bibinfo {author} {\bibfnamefont {V.~S.}\ \bibnamefont
  {Malinovsky}},\ }\href@noop {} {\emph {\bibinfo {title} {Principles of Laser
  Spectroscopy and Quantum Optics}}}\ (\bibinfo  {publisher} {Princeton Univ.
  Press, Princeton, NJ, USA},\ \bibinfo {year} {2011})\BibitemShut {NoStop}%
\bibitem [{\citenamefont {Boyd}(2003)}]{BoydNLOptic}%
  \BibitemOpen
  \bibfield  {author} {\bibinfo {author} {\bibfnamefont {R.}~\bibnamefont
  {Boyd}},\ }\href@noop {} {\emph {\bibinfo {title} {Nonlinear Optics}}}\
  (\bibinfo  {publisher} {Academic Press, San Diego, CA, USA},\ \bibinfo {year}
  {2003})\BibitemShut {NoStop}%
\bibitem [{\citenamefont {Wu}\ \emph {et~al.}(2009)\citenamefont {Wu},
  \citenamefont {Lin}, \citenamefont {Huang},\ and\ \citenamefont
  {Singh}}]{WuLinHuangEtAl2009}%
  \BibitemOpen
  \bibfield  {author} {\bibinfo {author} {\bibfnamefont {Y.-R.}\ \bibnamefont
  {Wu}}, \bibinfo {author} {\bibfnamefont {Y.-Y.}\ \bibnamefont {Lin}},
  \bibinfo {author} {\bibfnamefont {H.-H.}\ \bibnamefont {Huang}}, \ and\
  \bibinfo {author} {\bibfnamefont {J.}~\bibnamefont {Singh}},\ }\href
  {\doibase 10.1063/1.3065274} {\bibfield  {journal} {\bibinfo  {journal} {J.
  Appl. Phys.}\ }\textbf {\bibinfo {volume} {105}},\ \bibinfo {eid} {013117}
  (\bibinfo {year} {2009}),\ 10.1063/1.3065274}\BibitemShut {NoStop}%
\bibitem [{\citenamefont {Miller}\ \emph {et~al.}(1984)\citenamefont {Miller},
  \citenamefont {Chemla}, \citenamefont {Damen}, \citenamefont {Gossard},
  \citenamefont {Wiegmann}, \citenamefont {Wood},\ and\ \citenamefont
  {Burrus}}]{MillerChemlaDamenEtAl1984}%
  \BibitemOpen
  \bibfield  {author} {\bibinfo {author} {\bibfnamefont {D.~A.~B.}\
  \bibnamefont {Miller}}, \bibinfo {author} {\bibfnamefont {D.~S.}\
  \bibnamefont {Chemla}}, \bibinfo {author} {\bibfnamefont {T.~C.}\
  \bibnamefont {Damen}}, \bibinfo {author} {\bibfnamefont {A.~C.}\ \bibnamefont
  {Gossard}}, \bibinfo {author} {\bibfnamefont {W.}~\bibnamefont {Wiegmann}},
  \bibinfo {author} {\bibfnamefont {T.~H.}\ \bibnamefont {Wood}}, \ and\
  \bibinfo {author} {\bibfnamefont {C.~A.}\ \bibnamefont {Burrus}},\ }\href
  {\doibase 10.1103/PhysRevLett.53.2173} {\bibfield  {journal} {\bibinfo
  {journal} {Phys. Rev. Lett.}\ }\textbf {\bibinfo {volume} {53}},\ \bibinfo
  {pages} {2173} (\bibinfo {year} {1984})}\BibitemShut {NoStop}%
\bibitem [{\citenamefont {Wang}\ \emph {et~al.}(1990)\citenamefont {Wang},
  \citenamefont {Suna}, \citenamefont {McHugh}, \citenamefont {Hilinski},
  \citenamefont {Lucas},\ and\ \citenamefont
  {Johnson}}]{WangSunaMcHughEtAl1990}%
  \BibitemOpen
  \bibfield  {author} {\bibinfo {author} {\bibfnamefont {Y.}~\bibnamefont
  {Wang}}, \bibinfo {author} {\bibfnamefont {A.}~\bibnamefont {Suna}}, \bibinfo
  {author} {\bibfnamefont {J.}~\bibnamefont {McHugh}}, \bibinfo {author}
  {\bibfnamefont {E.~F.}\ \bibnamefont {Hilinski}}, \bibinfo {author}
  {\bibfnamefont {P.~A.}\ \bibnamefont {Lucas}}, \ and\ \bibinfo {author}
  {\bibfnamefont {R.~D.}\ \bibnamefont {Johnson}},\ }\href {\doibase
  10.1063/1.458280} {\bibfield  {journal} {\bibinfo  {journal} {J. Chem.
  Phys.}\ }\textbf {\bibinfo {volume} {92}},\ \bibinfo {pages} {6927} (\bibinfo
  {year} {1990})}\BibitemShut {NoStop}%
\bibitem [{\citenamefont {Hilinski}\ \emph {et~al.}(1988)\citenamefont
  {Hilinski}, \citenamefont {Lucas},\ and\ \citenamefont
  {Wang}}]{HilinskiLucasWang1988}%
  \BibitemOpen
  \bibfield  {author} {\bibinfo {author} {\bibfnamefont {E.~F.}\ \bibnamefont
  {Hilinski}}, \bibinfo {author} {\bibfnamefont {P.~A.}\ \bibnamefont {Lucas}},
  \ and\ \bibinfo {author} {\bibfnamefont {Y.}~\bibnamefont {Wang}},\ }\href
  {\doibase 10.1063/1.454913} {\bibfield  {journal} {\bibinfo  {journal} {J.
  Chem. Phys.}\ }\textbf {\bibinfo {volume} {89}},\ \bibinfo {pages} {3435}
  (\bibinfo {year} {1988})}\BibitemShut {NoStop}%
\bibitem [{\citenamefont {Haug}\ and\ \citenamefont
  {Schmitt-Rink}(1984)}]{HaugSchmitt-Rink1984}%
  \BibitemOpen
  \bibfield  {author} {\bibinfo {author} {\bibfnamefont {H.}~\bibnamefont
  {Haug}}\ and\ \bibinfo {author} {\bibfnamefont {S.}~\bibnamefont
  {Schmitt-Rink}},\ }\href {\doibase 10.1016/0079-6727(84)90026-0} {\bibfield
  {journal} {\bibinfo  {journal} {Prog. Quant. Electr.}\ }\textbf {\bibinfo
  {volume} {9}},\ \bibinfo {pages} {3 } (\bibinfo {year} {1984})}\BibitemShut
  {NoStop}%
\bibitem [{\citenamefont {Chemla}\ and\ \citenamefont
  {Miller}(1985)}]{ChemlaMiller1985}%
  \BibitemOpen
  \bibfield  {author} {\bibinfo {author} {\bibfnamefont {D.~S.}\ \bibnamefont
  {Chemla}}\ and\ \bibinfo {author} {\bibfnamefont {D.~A.~B.}\ \bibnamefont
  {Miller}},\ }\href {\doibase 10.1364/JOSAB.2.001155} {\bibfield  {journal}
  {\bibinfo  {journal} {J. Opt. Soc. Am. B}\ }\textbf {\bibinfo {volume} {2}},\
  \bibinfo {pages} {1155} (\bibinfo {year} {1985})}\BibitemShut {NoStop}%
\end{thebibliography}%
\end{document}